# Classical harmonic oscillator with quantum energy spectrum

Sergey A. Rashkovskiy

Institute for Problems in Mechanics, Russian Academy of Sciences,

Vernadskogo Ave. 101/1, Moscow, 119526, Russia, rash@hotbox.ru, rash@ipmnet.ru

**Abstract**

The classical dynamical system possessing a quantum spectrum of energy and "quantum" behavior is suggested and investigated. The proposed model can be considered as a dynamical variant of the old quantum theory for harmonic oscillator in which the Bohr-Sommerfeld quantization rule is absent and “quantum-mechanical” properties are the result of system behavior itself. For this dynamical system the classical model of Franck-Hertz experiment which allows explaining the experimentally observed regularities is suggested and investigated. The examples of calculations of Franck-Hertz experiment of within the limits of the suggested model of classical dynamical system are presented.

## I. INTRODUCTION

Inability of classical physics to explain the stability of atoms and the spectra of their radiation was one of the reasons of the development of quantum mechanics.

N. Bohr first attempted to answer these questions in his semi-classical theory of hydrogen-like atoms [1]. Bohr postulated, that atoms have discrete stationary energy levels, being on which, accelerated electron, in contradiction with classical electrodynamics, does not radiate the electromagnetic waves; the radiation occurs only during transition from one stationary energy level to another in the form of discrete energy amount – quanta.

The Bohr model was thereafter generalized for any periodic or multiperiodic motions by A. Sommerfeld [2] and W. Wilson [3]. The theory developed by them is known, as the old quantum mechanics. For these systems the Bohr-Sommerfeld quantization rule [4] allowed calculating the stationary energy levels which really corresponded to available experimental data of atoms spectrum.

The old quantum mechanics described the electron motion in atom within the limits of the classical mechanics, however it stated that not all kinds of motions are allowed but only those motions which satisfy the Bohr-Sommerfeld quantization rules. Thus the Bohr-Sommerfeld quantization rules, actually, were the rules of selection of true stationary states from all possible electron motions predicted by classical mechanics. All other stationary classical trajectories of electron, which do not satisfy the Bohr-Sommerfeld rules of selection were rejected and considered as impossible in the old quantum mechanics.

A weakness of the old quantum theory was its inconsistency. On one hand the electron motion in atom remained a classical one, but on the other hand the new rule of artificial selection of motions was entered which did not have an explanation within the limits of the classical mechanics.

It is necessary to note, that this rule of selection did not have any physical substantiation, except the fact that results of the theory described available experimental data on a spectrum of hydrogen-like atoms correctly.

Inability of classical physics to eliminate internal contradictions of the Bohr-Sommerfeld theory has forced the physicists to be engaged in developing of new theory which finally became the modern quantum mechanics and the quantum field theory.

In the course of development of quantum mechanics the representations was formed [4] that it is impossible to construct the classical mechanical system which would possess by discrete energy levels and "correct quantum behavior", i.e. which would be able to undergo a transition spontaneously from one stationary level to another, radiating discrete amount of energy - quanta.

In the present work we will consider and investigate the classical mechanical system which has stationary states with discrete (quantized) energy levels. The system is able to stay in these states for an infinitely long time when external disturbances are absent, but at the same time the system undergoes a transition spontaneously to a lower energy level under the action of even infinitesimal disturbances, losing the energy in the form of discrete portions.

By the example of harmonic oscillator, we will show how the dynamical variant of the old quantum theory, which is free from above mentioned contradiction of Bohr-Sommerfeld theory can be built. In this model the stationary levels and transitions between them do not arise as a result of logical operation (Bohr-Sommerfeld quantization rule) but as natural result of behavior of nonlinear dynamical system.

It is important to note that the purpose of this work is not "construction of the classical theory of quantum systems", but more narrowly – the demonstration of existence of classical systems with "correct" quantum behavior.

## II. MODEL OF DYNAMICAL SYSTEM

Let us briefly consider the well-known peculiarities of behavior of quantum systems from point of view of theory of dynamical systems.

It is well-known, the quantum systems, including atoms, harmonic oscillator etc., have a discrete set of stationary states with discrete energy levels. The stationary state with the least energy is the ground state; all other stationary states are excited. Being in one of such states, the system has a constant energy and it is able to stay in such a state for an infinitely long time when the external disturbances are absent. It is impossible to disturb the system from the ground state even under the enough strong action (the Franck-Hertz experiment [5] is an example of such a statement). Thus, it can be stated that the ground state is absolutely stable. If the system is in an excited state it can undergo a transition spontaneously to a lower energy level. This transition is accompanied by emission of the energy in the form of quantum. A simple mechanical interpretation can be given to this process: the excited state is absolutely unstable and the system undergoes a transition to a lower energy level under the action even small disturbances which are always present in space in the form of waves and fields. An isolated quantum system (i.e. a system which is acted on by only small disturbances) always reverts to the ground state having radiated all excess of energy. It can remain in this state for unlimited long time.

Spontaneous transition of quantum system to a stationary state with higher energy level is impossible. Such transition can occur only forcedly as a result of external action with finite energy. During such transition the discrete amount of energy (quantum) equal to difference of energy of those levels between which the transition occurs should be supplied to the system. In other words, the transition of quantum system to higher energy level is possible only as a result of absorption of the corresponding energy quanta by the system.

Using these representations, we can build a classical dynamical system the behavior of which is similar to the behavior of quantum harmonic oscillator.

It has to be noted that the classical dynamical system is understood here as a point-like particle the state of which is completely defined by its coordinate and velocity at a given time. Moreover only classical forces (i.e. local forces, without delay etc.) act on this system.

Let us list the expected properties of such a dynamical system.

1. The energy of harmonic oscillator at any time is defined by classical expression

$$E = \frac{m}{2}\left(\dot{q}^2 + \omega^2 q^2\right) \tag{1}$$

where $q$ is classical coordinate of the particle, $m$ is its mass, $\omega$ is frequency.

2. Harmonic oscillator can exist in a certain number of stationary states with the energies

$$E_n = \hbar\omega\left(n + \frac{1}{2}\right) \tag{2}$$

where $n$ =0,1,2,…

The states corresponding to other values of energy are nonstationary.

The harmonic oscillator is a conservative system only in stationary states with energies (2) and its energy remains constant during motion.

At any other energies which do not coincide with energy levels (2) the oscillator is not a conservative system and during its motion the energy loss occurs; this process can be interpreted as an emission of the energy by the system.

3. Being in a stationary state with energy (2) the harmonic oscillator is fully classical one in the sense that it is described by linear equation of harmonic oscillations

$$\ddot{q} + \omega^2 q = 0 \tag{3}$$

4. The ground state ($n$=0) is absolutely stable. The oscillator can exist for infinitely long time in this state even under external actions with finite energy.

5. The excited states ($n \neq 0$) are absolutely unstable. Even for small deviations of energy from level (2) to the lower energies (i.e. if a small amount of energy is taken away from the system) the system will undergo a transition to a lower stationary energy level (2) and lose a part of its energy (quantum) in the process. This process corresponds to emission of the energy quantum equal to the difference of energy of those levels between which transition occurs. Under the action of infinitesimal disturbances which always exist in the surrounding medium this process can be considered as a spontaneous transition (spontaneous emission); under the action of disturbances with finite energy such transition can be considered as a stimulated transition (stimulated emission).

6. In the case there are deviations from excited states (2) towards higher energy (e.g. the system absorbs small amount of energy $\Delta E < E_{n+1} - E_n$) the system has to either come back to the initial energy level with emission of earlier received energy $\Delta E$ or undergo a transition to a lower energy level with emission of corresponding energy quanta. The last case can also be interpreted as stimulated transition (stimulated emission).

7. For transition to a higher excited level (2) the system has to absorb a finite amount of energy not less than the difference of energy levels between which a transition will occur; i.e. for transition from the energy level $n$ to the energy level $k > n$ the system should absorb the energy $\Delta E \geq E_k - E_n$. This process corresponds to stimulated transition.

In accordance with properties 4-7, if such a system is placed into real space, filled with fields and radiation, but isolated from disturbances of finite energy, the excited states with $n \neq 0$ will have a finite lifetime, while in the ground state $n = 0$ the system can exist for an infinitely long time.

The indicated properties of the dynamical system mean that generally its energy should not be a constant. The equation of energy changing can be written as follows

$$\frac{dE}{dt} = -f(q, \dot{q}, E) \tag{4}$$

where $f(q, \dot{q}, E)$ is some function satisfying the following conditions.

1. In stationary states, which are the energy levels (2)

$$f(q, \dot{q}, E) = 0 \tag{5}$$

and discrete energy levels (2) of the harmonic oscillator should be solutions of the algebraic equation (5).

2. Outside of the stationary levels (2) it should be

$$f(q, \dot{q}, E) > 0 \text{ for } E > \frac{1}{2}\hbar\omega \quad \text{and} \quad f(q, \dot{q}, E) < 0 \text{ for } E > \frac{1}{2}\hbar\omega \tag{6}$$

The simplest function $f(q, \dot{q}, E)$ satisfying these conditions has the form

$$f(q, \dot{q}, E) = ma_0\left(E - \frac{\hbar\omega}{2}\right)\dot{q}^2 \cos^2(\pi E/\hbar\omega) \tag{7}$$

where $a_0 > 0$ is some function. Hereinafter we consider $a_0 = const$.

The graph of function (7) is shown at Fig. 1; the arrows indicate the direction of system motion along energy axis.

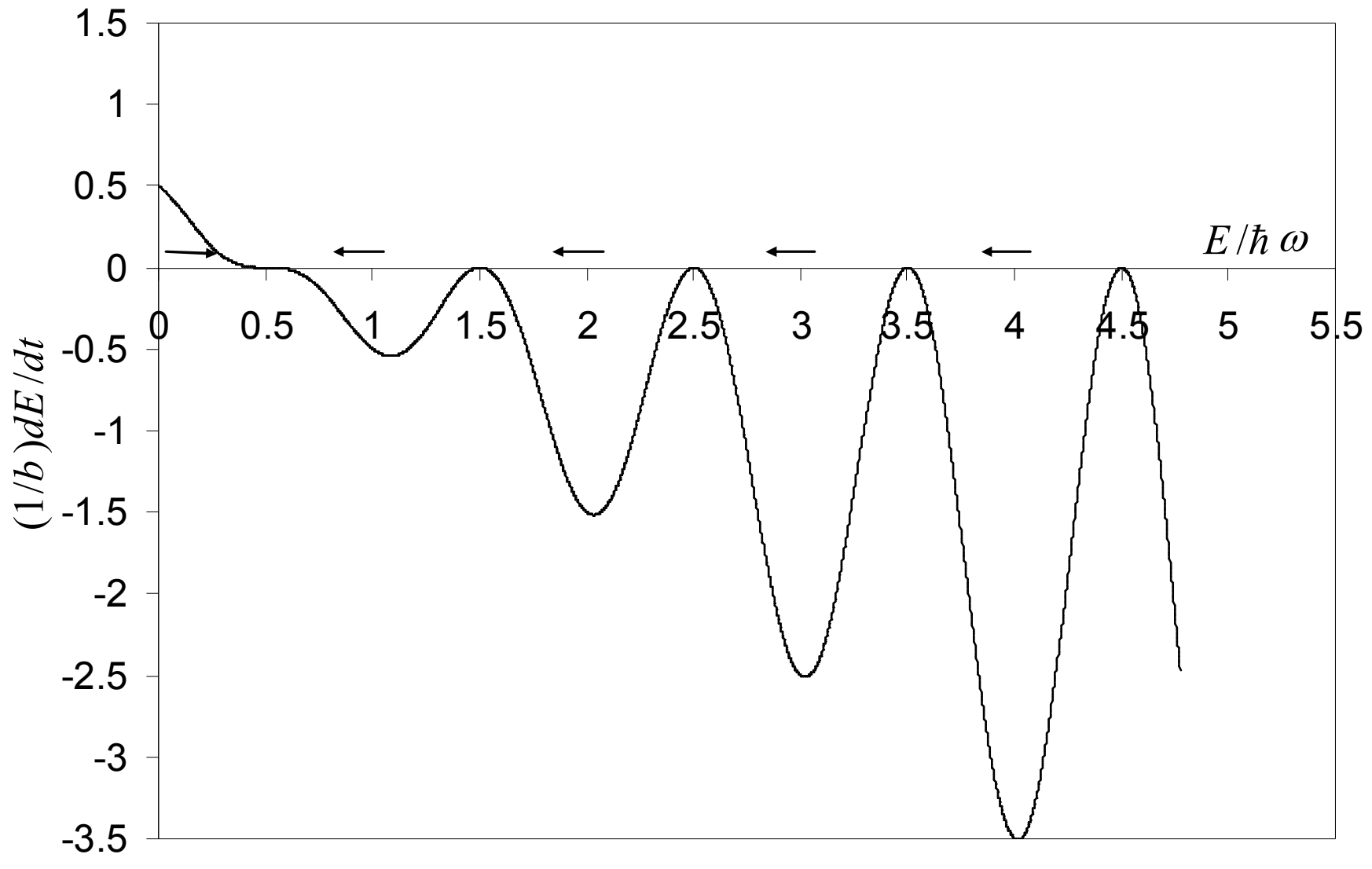

FIG. 1. Function (4), (7); $b = ma_0\dot{q}^2\hbar\omega \geq 0$

Taking into account (1), (4), (7), the motion equation can be presented as

$$\ddot{q} + \omega^2 q = -a_0\left(E - \frac{\hbar\omega}{2}\right)\dot{q}\cos^2(\pi E/\hbar\omega) \tag{8}$$

The system (8) is a classical harmonic oscillator which is under the action of the additional force

$$F_q = -ma_0\left(E - \frac{\hbar\omega}{2}\right)\dot{q}\cos^2(\pi E/\hbar\omega) \tag{9}$$

On stationary levels (2) this force is equal to zero and particle motion at such values of energy is conservative . This motion is described by classical linear equation of harmonic oscillations (3). Outside of the stationary levels (2) the force (9) can be interpreted as a "frictional force" $F_q = -\mu\dot{q}$ with variable coefficient of friction $\mu = ma_0\left(E - \frac{\hbar\omega}{2}\right)\cos^2(\pi E/\hbar\omega)$ which appears, e.g. between the particle and its own radiation (its electromagnetic field). This force can be called the "quantum friction".

## III. RESULTS AND DISCUSSION

The equation (8) has been solved numerically.

For convenience of numerical investigations the equation (8) is reduced to non-dimensional form.

Leaving former symbols for non-dimensional variables, the following substitution can be made

$$\omega t \to t,\ (m\omega/\hbar)^{1/2} q \to q,\ E/\hbar\omega \to E \tag{10}$$

In non-dimensional variables the energy (1) is defined by expression

$$E = \frac{1}{2}\left(\dot{q}^2 + q^2\right) \tag{11}$$

and equation (9) takes the form

$$\ddot{q} + q = -\alpha\dot{q}\left(\dot{q}^2 + q^2 - 1\right)\cos^2\left(\frac{\pi}{2}(\dot{q}^2 + q^2)\right) \tag{12}$$

where the positive constant $\alpha$ is defined by expression

$$\alpha = \frac{1}{2} a_0 \hbar \tag{13}$$

and it is the single indeterminate parameter of the model.

The equation (12) was calculated for different initial conditions $q(0) = q_0,\ \dot{q}(0) = V_0$.

Calculations show that the system eventually undergoes a transition into one of the stationary states (2) at any initial conditions.

The values of energy which system reaches with $\alpha$=0.1 by the time $t$=100 at different initial velocity $V_0$ are shown at Fig.2. We assumed $q_0 = 0$ in these calculations.

This figure shows that a multitude of initial values of velocity $V_0$ correspond to the same stationary state. This means that the particle can reach the same stationary energy level having the different initial velocity. We can see that at high values of quantum numbers $n >> 1$ the energy of oscillator is approximately defined by the classical expression

$$E = \frac{1}{2} V_0^2 \tag{14}$$

Moreover the higher $n$ the more exactly this formula describes the results of calculations on the model (12). This result can be considered as a manifestation of Bohr's correspondence principle: a quantum system at high quantum numbers behaves as a classical system [4].

The system reaches the ground state ($n$=0) very slowly asymptotically because in its vicinity $\varepsilon = E - \frac{1}{2} << 1$, $\frac{d\varepsilon}{dt} \sim -\varepsilon^3$, i.e. $\varepsilon \sim t^{-1/2}$. Calculations show that system with $\alpha$=0.1 has not enough time to reach the ground state during $t = 100$. Therefore the data represented on the Fig.2 for the ground state is slightly different from $E = \frac{1}{2}$. The character of tendency to the ground state and the changes of the system energy in this process are shown at Fig. 3-6.

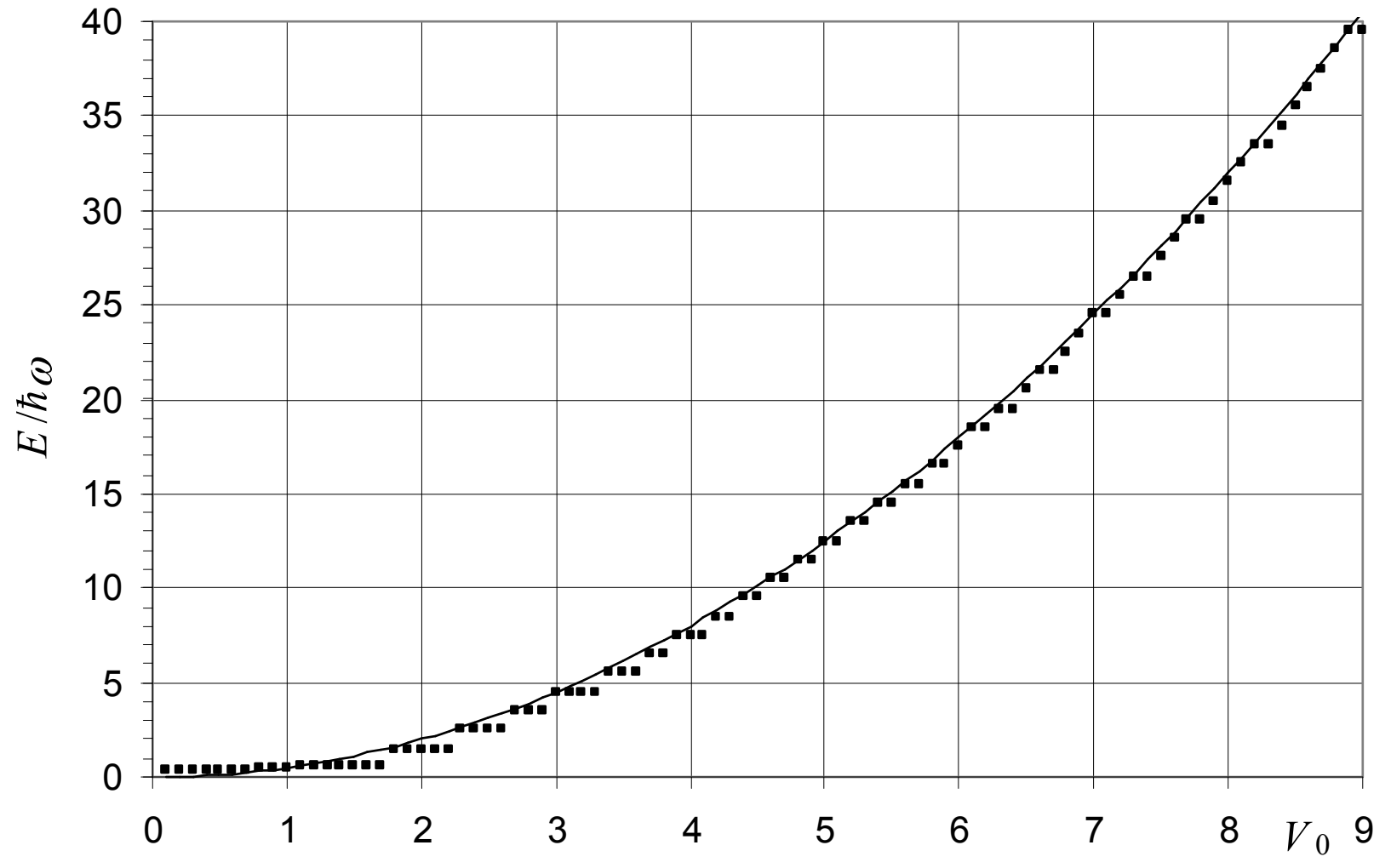

FIG. 2. Dependence of system energy at $t$=100 on initial velocity: the markers are the solutions of the equation (12) for $\alpha$=0.1, the line is the classical dependence (14)

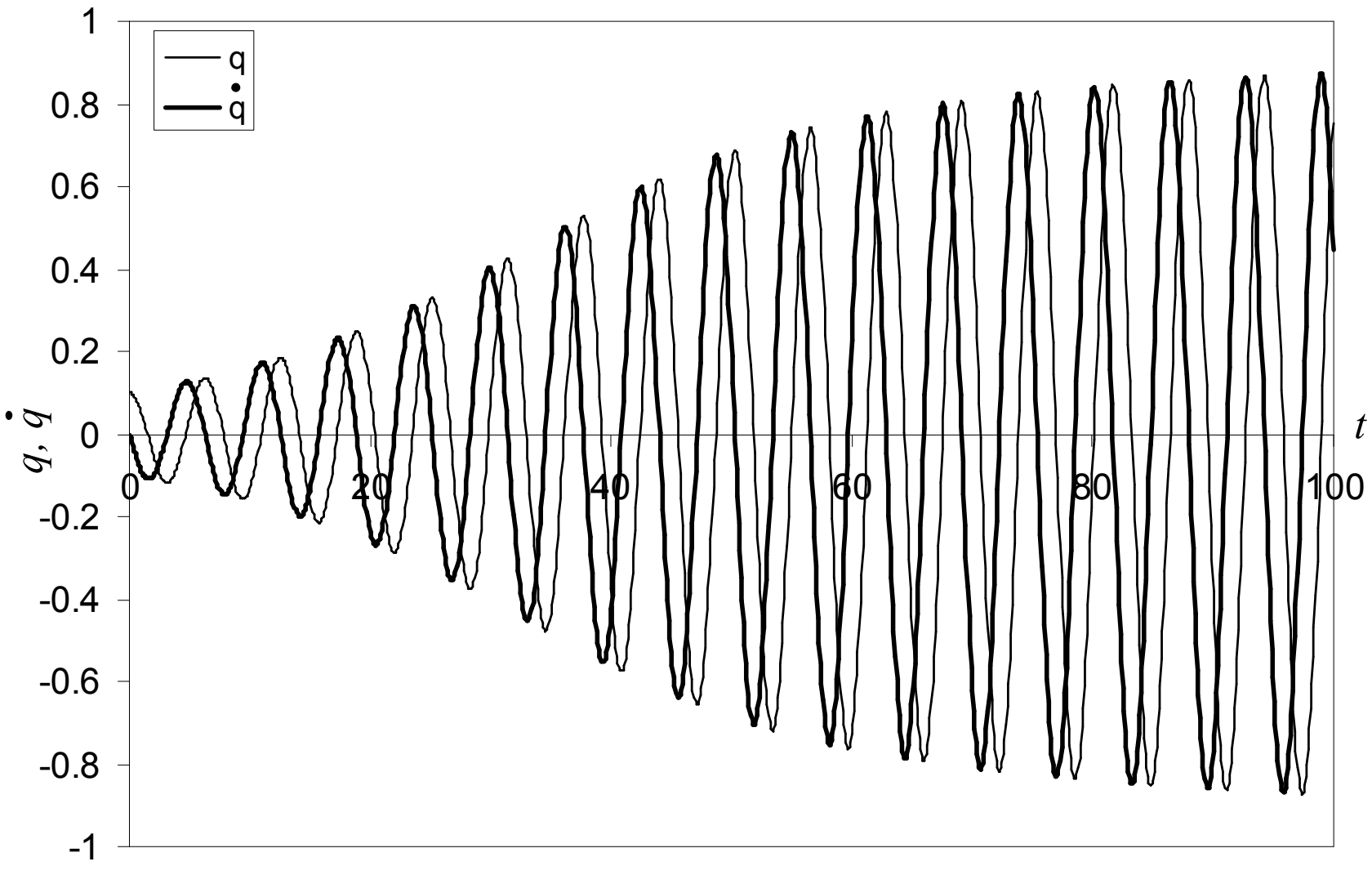

FIG. 3. Transition of the system to the ground state for $q_0$=0.1, $V_0$=0, $\alpha$=0.1.

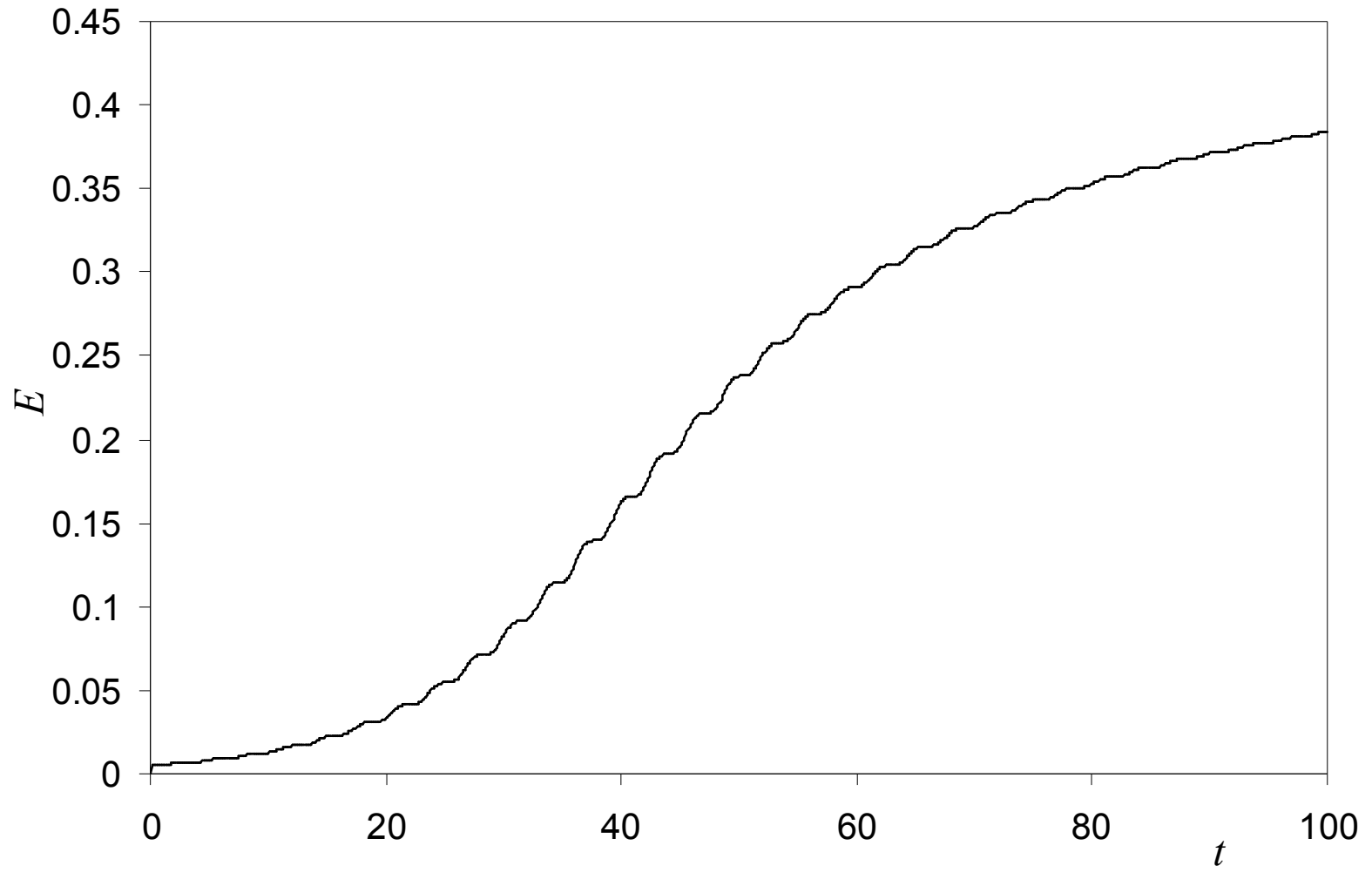

FIG. 4. Changing of non-dimensional energy of the system during its transition to the ground state; $q_0$ =0.1, $V_0$ =0, $\alpha$ =0.1.

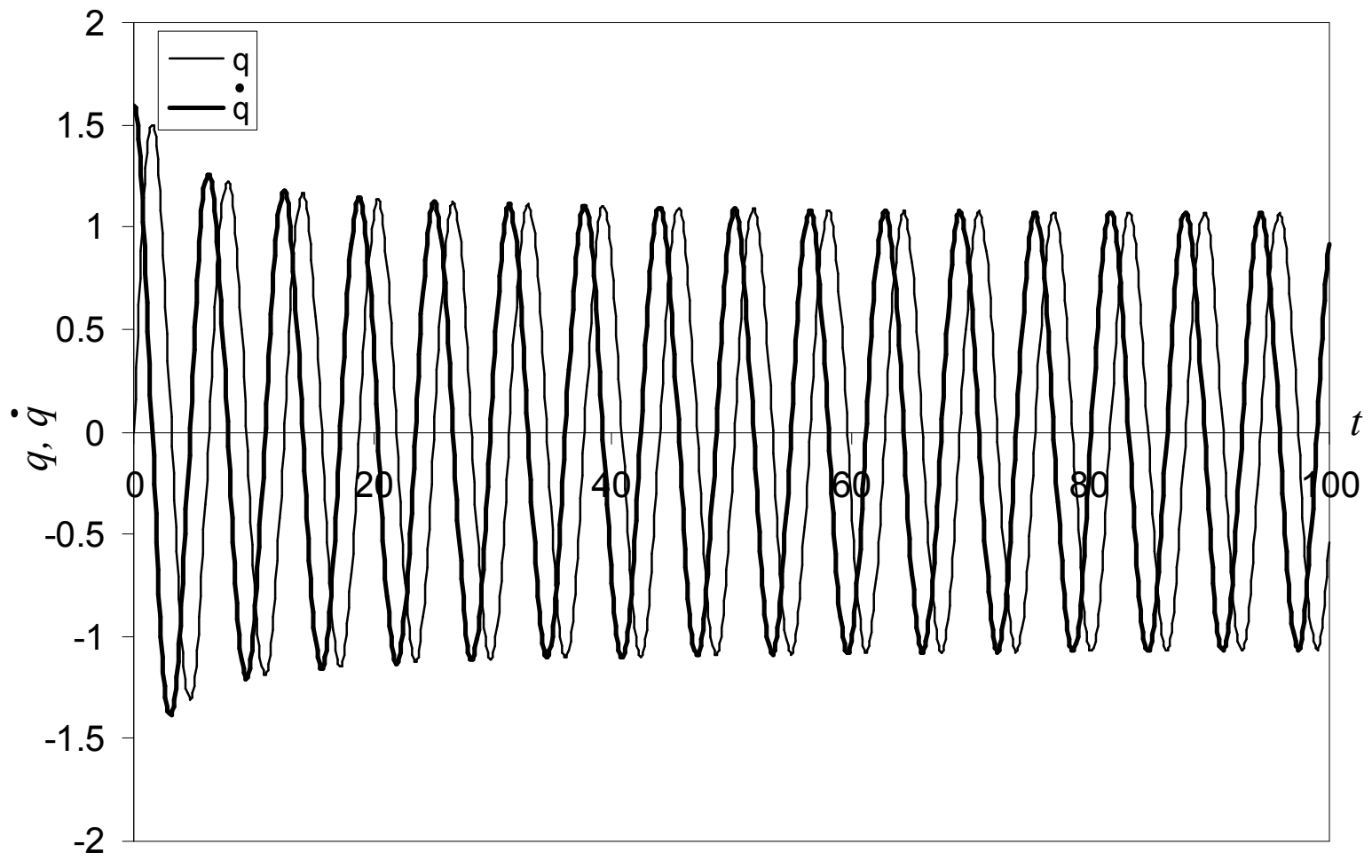

FIG. 5. Transition of the system to the ground state for $q_0$ =0, $V_0$ =1.6, $\alpha$ =0.1

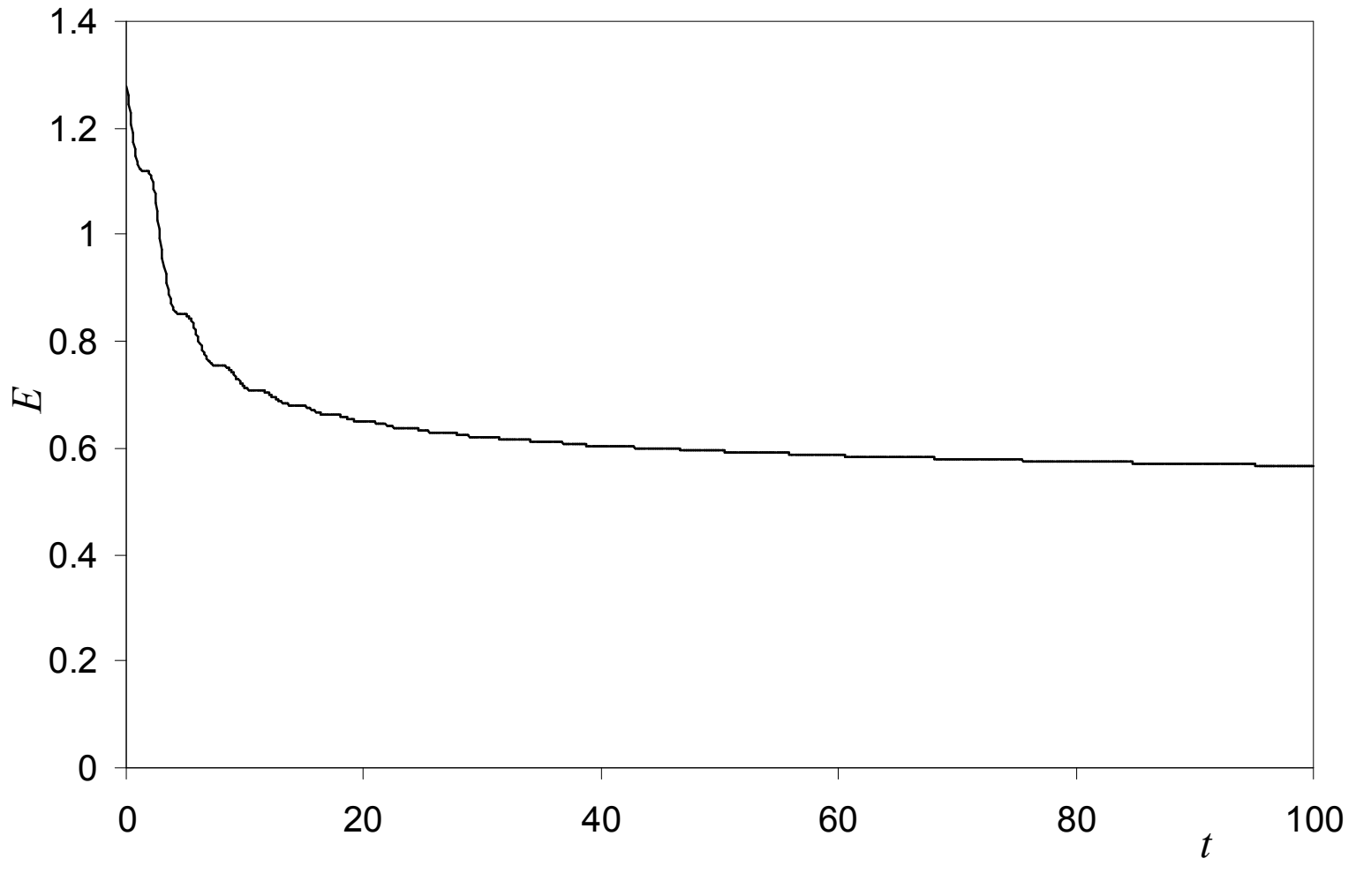

FIG. 6. Changing of non-dimensional energy of the system during its transition to the ground state; $q_0$ =0, $V_0$ =1.6, $\alpha$ =0.1.

A long supervision of the system existing in one of the excited states shows that in some time it undergo a transition spontaneously to a lower energy level; moreover the transitions occur more intense with the increase of parameter $\alpha$ .

The "occupancy" of the energy levels for $\alpha = 10$ at the time $t$=100 as a function of initial velocity $V_0$ is shown at Fig.7. We assumed $q_0 = 0$ in the calculations.

The observations of behavior of the model system show that a peculiar "fall of the levels" occurs: the system existing in the excited states eventually undergoes a transition to a lower energy level. As a result of this process the excited states with higher energy are devastated while the lower energy levels are filled up.

It has to be noted that the stationary energy levels remain quantized even in this case, however a one-to-one correspondence between a system energy at given time and an initial velocity is absent.

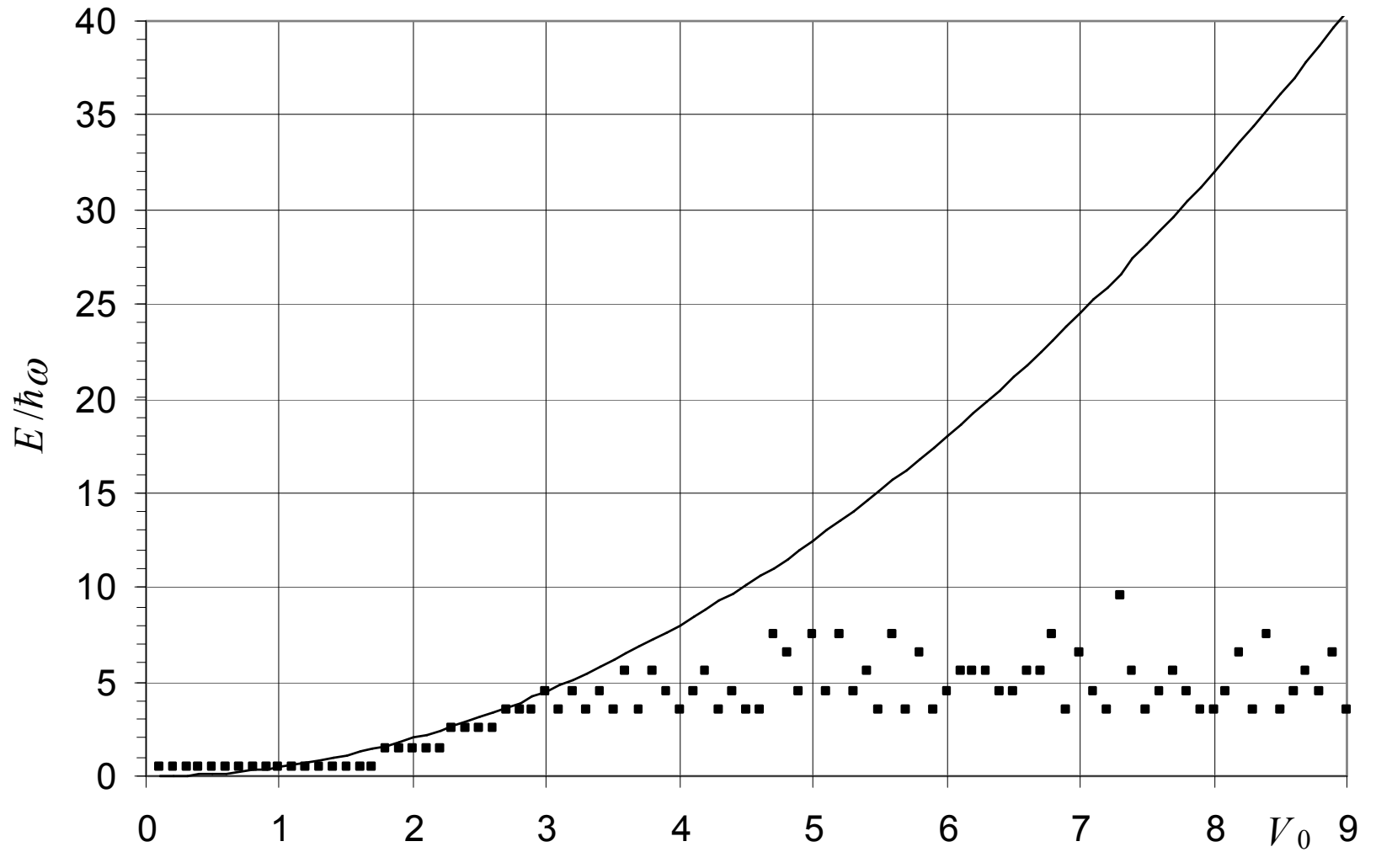

FIG. 7. "Occupancy" of quantum levels at $t$=100 for different initial velocities: the markers are solutions of the equation (12) for $\alpha$ =10, the line is the classical dependence (14)

Calculations show the lower energy level, the longer the system remains on it: lower energy levels are destroyed more slowly than higher ones. However at $t \to \infty$ the system always undergoes a transition to the ground state, independently of initial conditions.

Such behavior of the system under consideration looks like a behavior of real atoms, where the occupancy of lower energy levels is always higher, than the occupancy of the higher levels, in so doing the quantum levels with the big quantum numbers are practically non-occupied.

The higher parameter $\alpha$ the faster a "fall of the levels" occurs. For example, at $\alpha$ =1 during time $t$=100 all higher energy levels "fall" beginning from $n \approx 16$. At lower quantum numbers all energy levels are stable and one-to-one correspond to initial velocity (initial energy). At small $\alpha$ the system is attracted only to closer energy level for given initial energy (Fig.2) and remains here for a long time.

The old quantum theory considered only the stationary states of atoms, while a process of transition from one stationary energy level to another was not considered. It was simply assumed, that at such transitions an emission of quanta occurs with energy equal to a difference of energy of those levels between which a transition occurs.

Within the limits of the model under consideration we can observe the process of spontaneous transition of the system from one stationary energy level to another. It allows seeing how the system loses the energy during such a transition.

The results of the calculations for relatively long time of the process are shown at Fig. 8 and 9. It shows that the system stays at one of the excited states for some time but eventually it undergoes a transition spontaneously to a level with a lower energy.

The system has a similar behavior and at other values of parameter $\alpha$ .

The process of spontaneous transition of the system between excited states is shown in detail at Fig. 10 and 11.

A drastic change of particle velocity occurs during the transition while the particle coordinate is changed weakly during the transition.

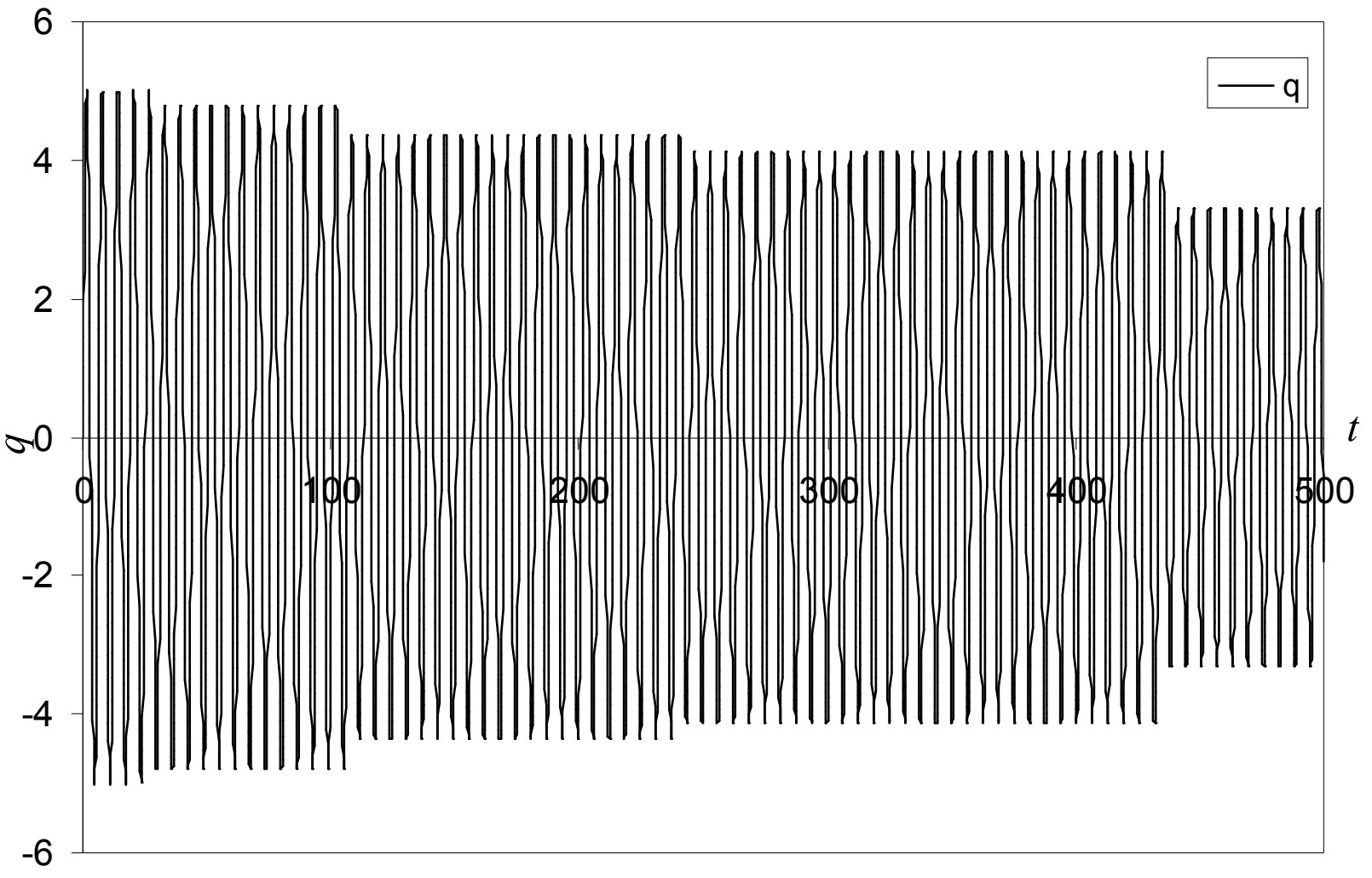

FIG. 8. Behavior of the system at $q_0$ =0, $V_0$ =5.0, $\alpha$ =7

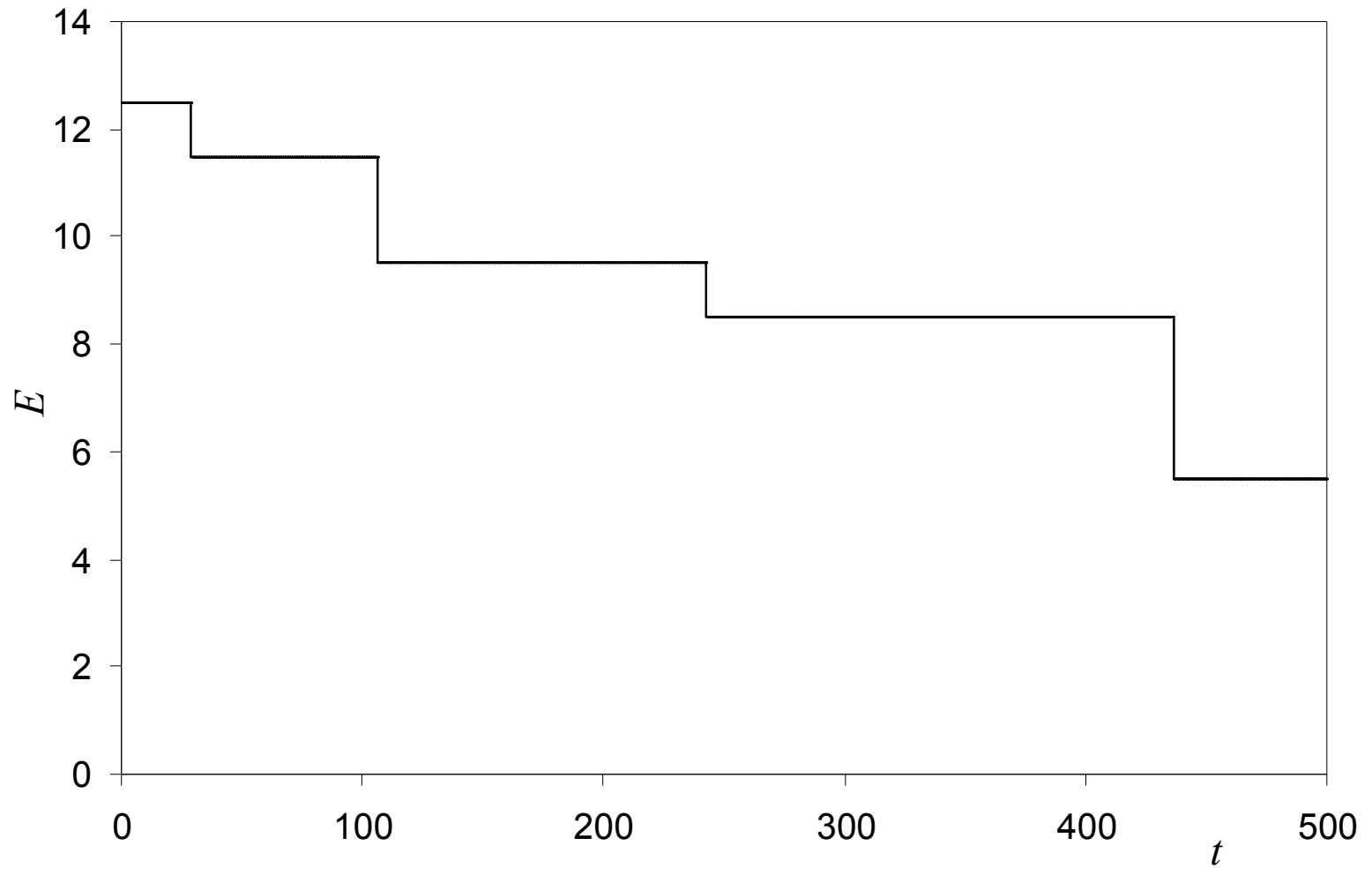

FIG. 9. Changing of non-dimensional energy of the system in the process for $q_0$ =0, $V_0$ =5.0, $\alpha$ =7. Energy levels on which the system was captured are (left to right) $n$ =12, 11, 9, 8, 5.

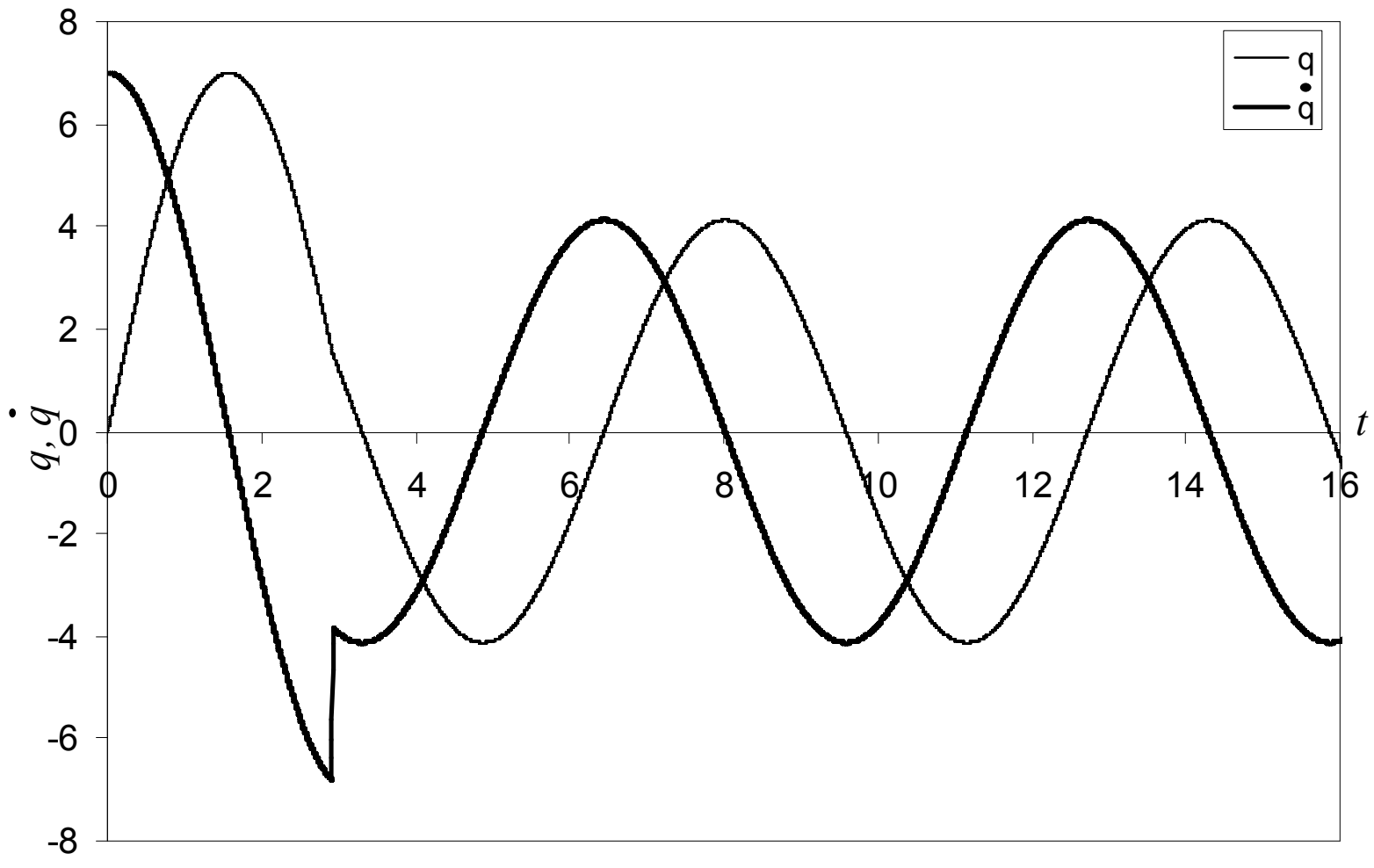

FIG. 10. Spontaneous transition in the system at $q_0$ =0, $V_0$ =7.0, $\alpha$ =10

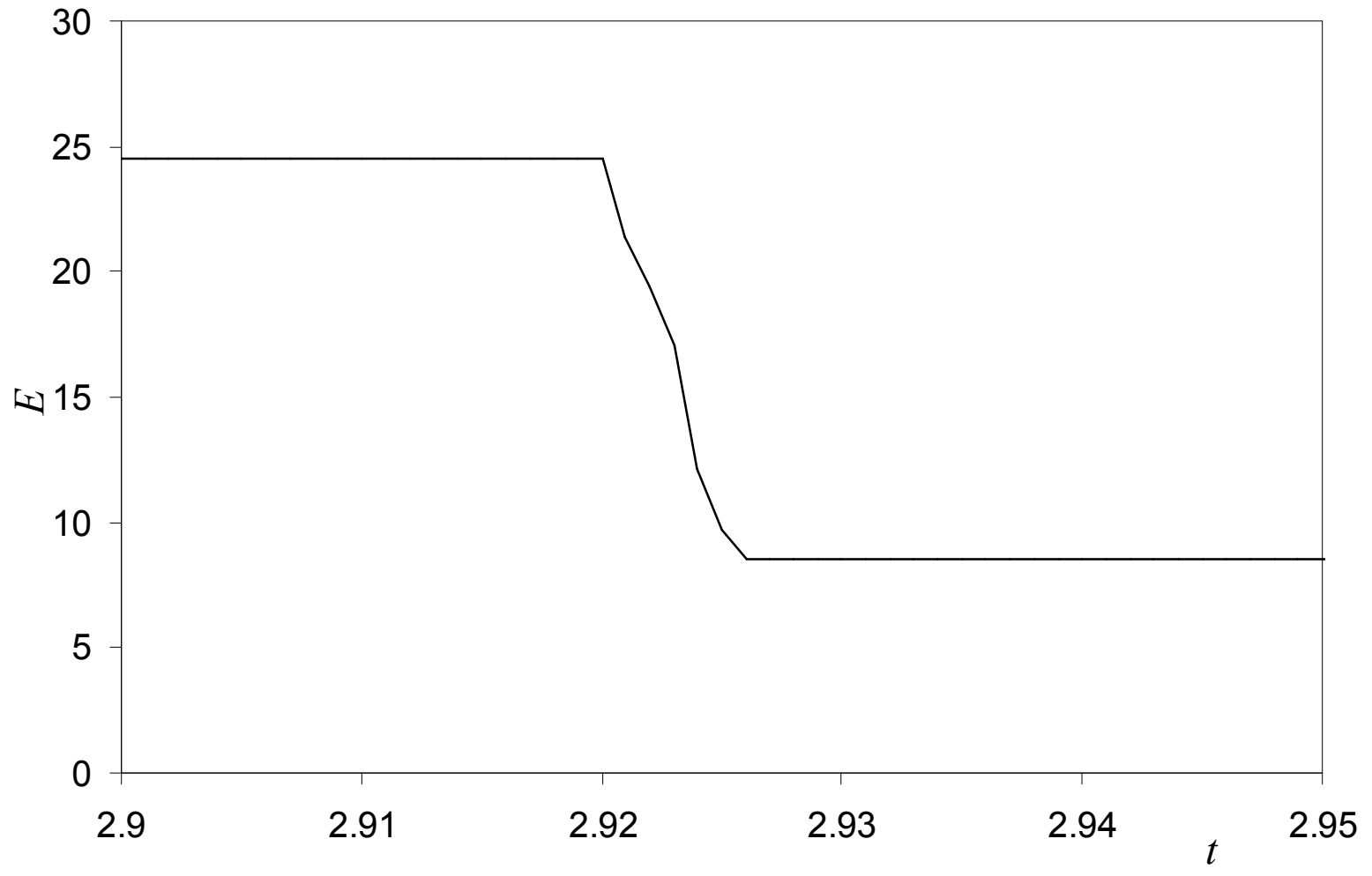

FIG. 11. Changing of non-dimensional energy of the system at spontaneous transition; $q_0$ =0, $V_0$ =7.0, $\alpha$ =10. Duration of transient period is ~0.006

It can be seen that a transition onto the new energy levels and consequently the loss (emission) of the energy occurs during the time essentially less than period of oscillations of the system, i.e. practically instantly.

It means, an energy emitted by the system is not "spread" on time, and represents a compact portion - quantum. It also correlates with representations about emission of quanta by atoms.

## IV. LIFETIME OF THE EXCITED STATES

Let us consider the lifetime of the system (12) in the excited states.

A changing of the system energy over a long period of time is shown at Fig. 12.

It shows that the lower energy level the longer its lifetime, this means that the lower energy levels are more stable than higher ones.

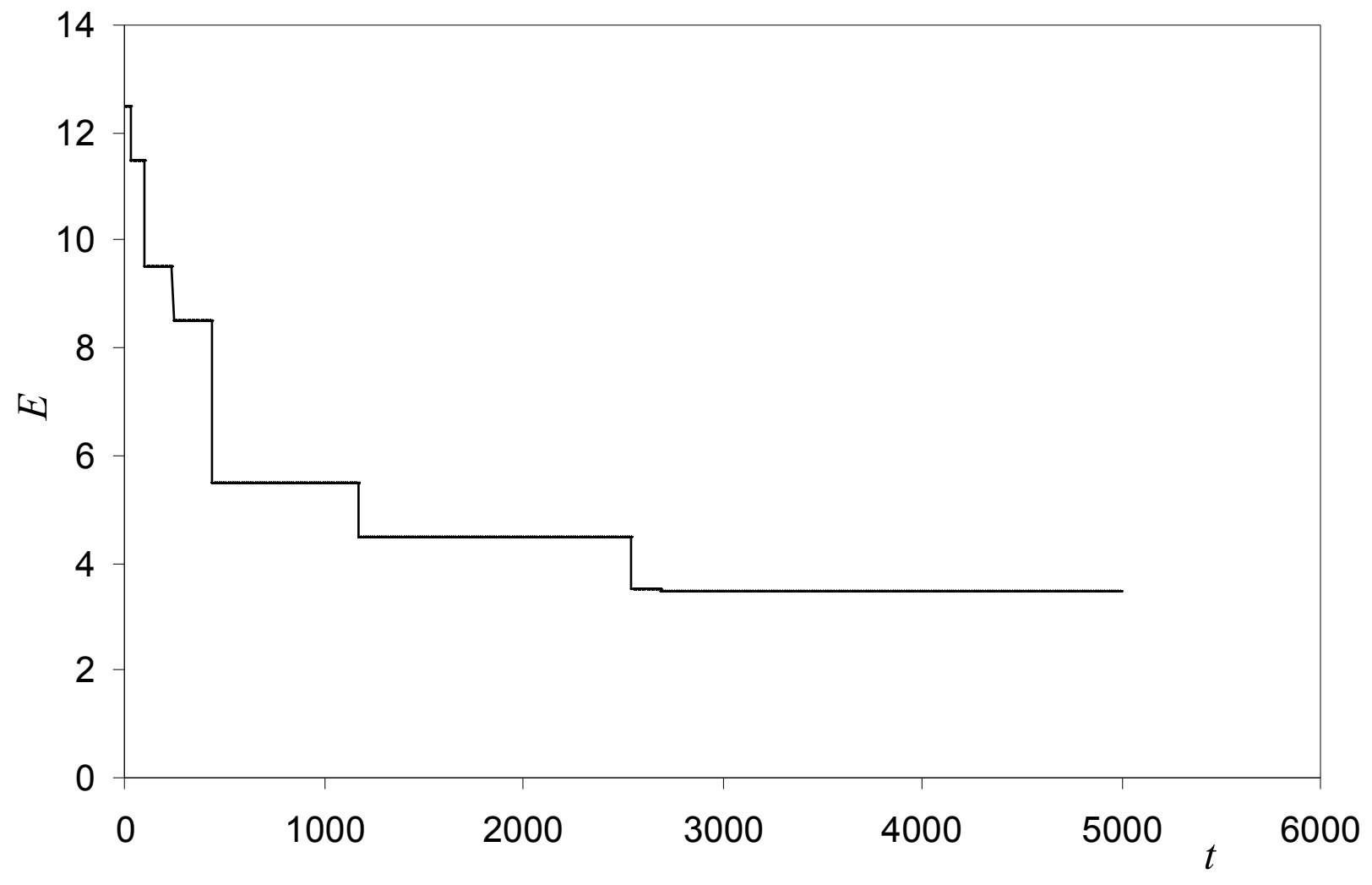

FIG. 12. Changing of non-dimensional energy of the system for $q_0$=0, $V_0$=5.0, $\alpha$=7

Calculations show that initial velocity $V_0$ does not effect practically the lifetime of excited states. This effect occurs only at the first level which the process is started from.

The dependence of the lifetime of the excited state on its energy is shown at Fig.13. This dependence can be approximated by the power function

$$\tau = AE^{-\beta} \tag{15}$$

where the parameter $A$ depends on $\alpha$ (see Fig. 14);

$$\beta=3.11\pm0.05, \quad A = A_0\alpha^{-1}, \quad \ln A_0=13.93 \tag{16}$$

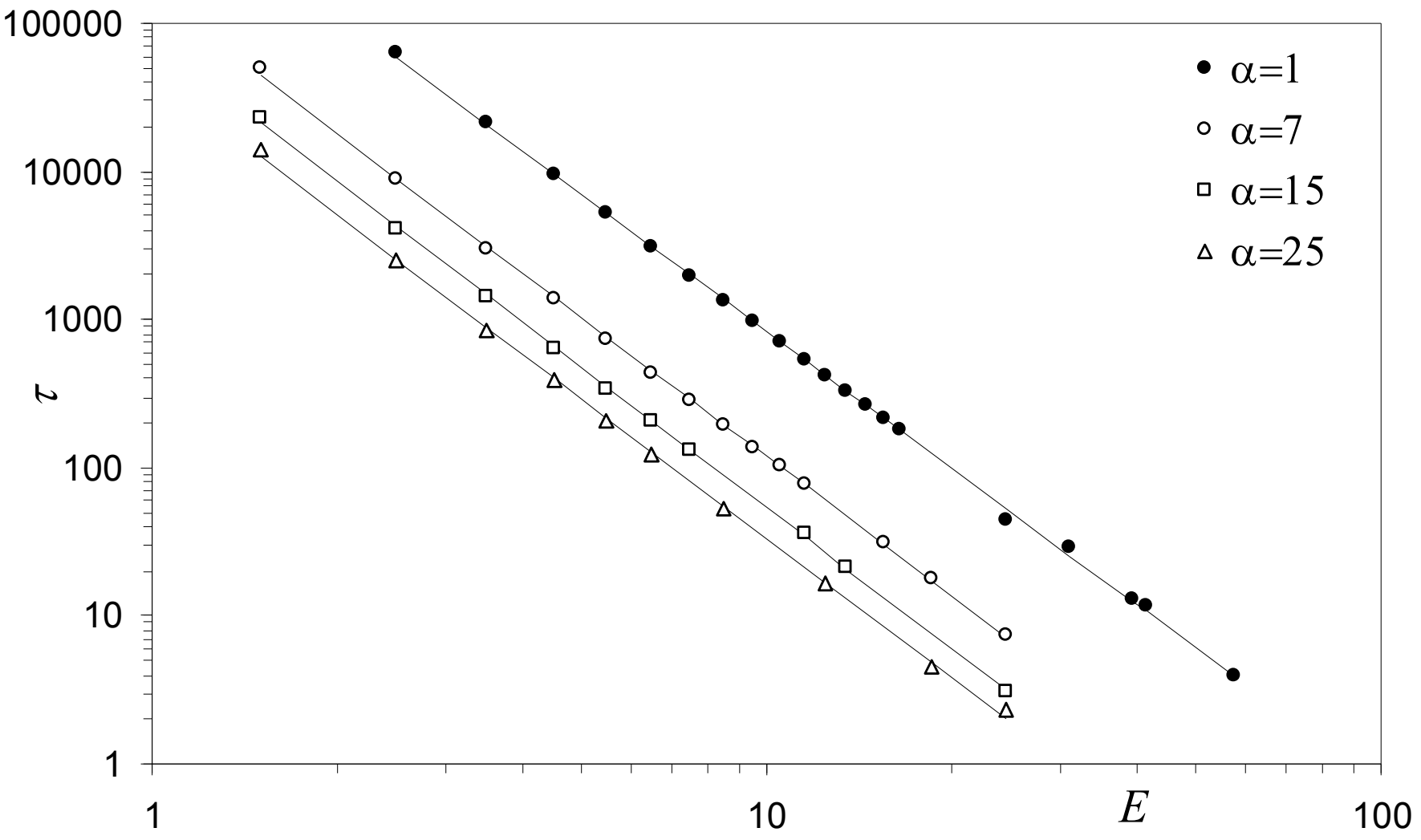

FIG. 13. Dependence of lifetime of excited state on its energy; the markers are the results of direct calculations for model (12), the lines are approximations (15), (16)

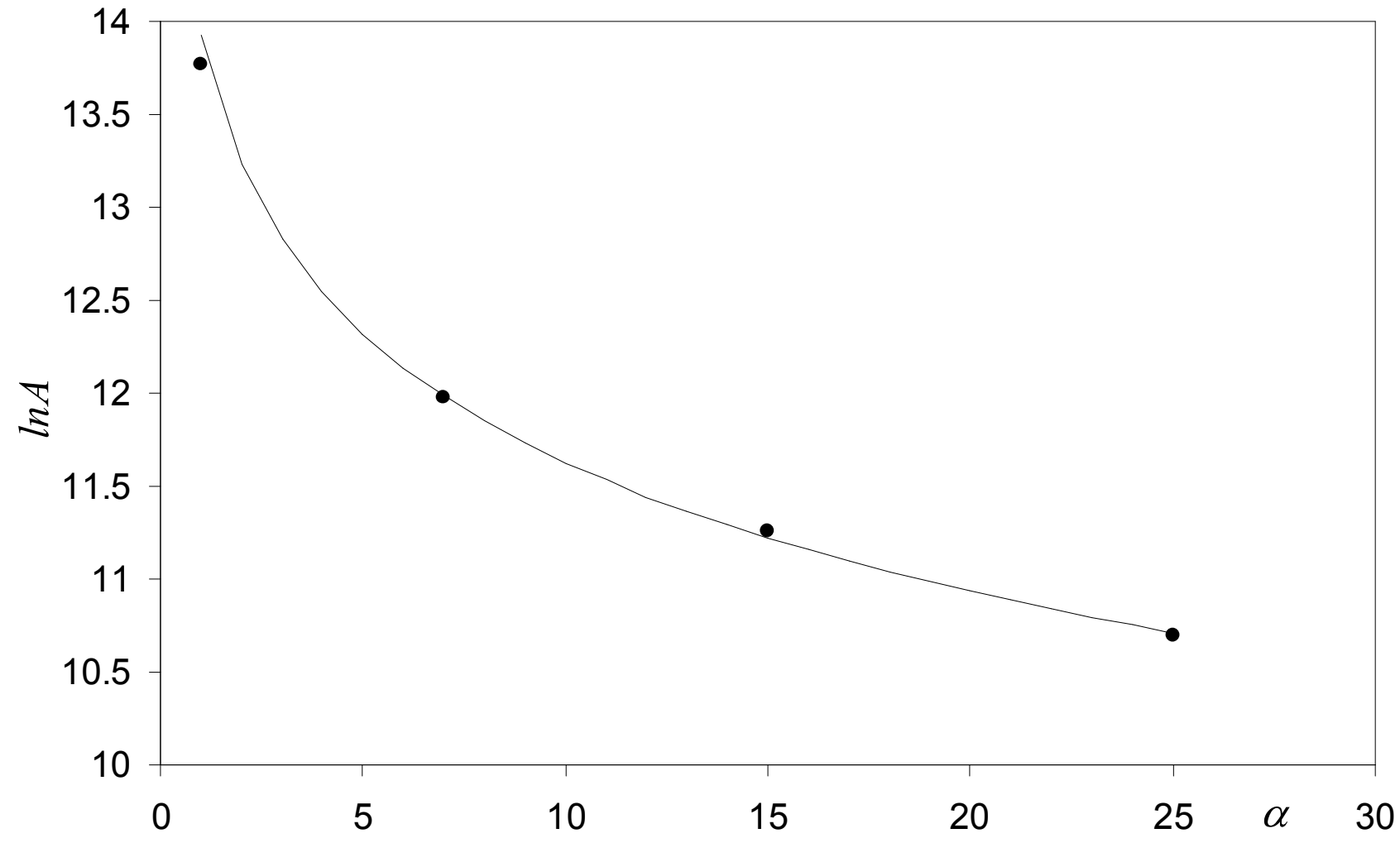

FIG. 14. Dependence of $\ln A$ on $\alpha$ ; dots correspond to different lines on Fig. 13; the line is approximation (16)

Let us discuss briefly the reason of spontaneous transition. As evident from Fig. 1 all excited states are absolutely unstable and even a slight disturbances remove the system from these stationary states. In our calculations such disturbances appear as a result of numerical solution of equation (12) with small but finite time step. It can be expected that if the system is under the action of small external disturbances the lifetime of excited states will be changed, and will depend on the characteristics of the disturbances. This problem represents a certain interest and it will be considered in future.

## V. FRANCK-HERTZ EXPERIMENT

Franck-Hertz experiment [5] on scattering of electron on atoms is one of the fundamental experiments in quantum physics.

This experiment confirmed Bohr's quantized model of the atom by demonstrating that atoms could indeed only absorb (and be excited) by discrete specific amounts of energy (quanta). It was the most important direct proof of quantization of energy levels in atoms.

At present it is widely believed that the Franck-Hertz experiment cannot be explained within the limits of the classical mechanics, because this is a pure quantum phenomenon [4].

Let us show that a “classical” explanation of Franck-Hertz experiments can be given within the limits of the classical model of oscillator with quantum energy spectrum. Moreover the process of scattering of free electron by the atom (in our case by the oscillator) can be calculated in detail (in classical sense), and the results of calculations really demonstrate the peculiarities observed in the Franck-Hertz experiment.

Let us consider a qualitative explanation of the Franck-Hertz experiments for the model of oscillator under consideration.

To avoid confusion, at the description of this process we will use two notions: collision and scattering. Collision means a direct mechanical collision of free electron with oscillator while a scattering means a final result of the process of electron-oscillator interaction as a whole.

Let us imagine that a free moving electron collides with the oscillator (in a real experiment with atom). For simplicity we will assume, that this collision is absolutely elastic. As a result of the collision the momentum and energy of particles are changed abruptly in accordance with usual laws of classical mechanics. Moreover, let us assume that this collision is instant: before and after the collision the oscillator and electron do not interact directly.

We start from the case where the electron energy is less than the difference of the energies between the first excited state and the ground state of oscillator. As a result of this collision the oscillator leaves the ground state but does not reach the first excited state. In accordance with dynamical equation (12) the oscillator tends to come back to the ground state under the action of “quantum friction”. Energy losses during this process can be considered as electromagnetic radiation which the oscillator emits when it comes back to the ground state. In the absence of free electron when disturbances in the oscillator were caused by other reasons, it can be assumed that energy emitted by oscillator is simply dissipated in environment. In case of interaction of oscillator with free electron we assume that the energy emitted by the oscillator in its motion to the ground state is absorbed by free electron. If it is assumed that all energy emitted by the oscillator is absorbed by free electron, then, eventually, free electron will return itself all the energy lost at collision with the oscillator and the scattering of electron will be elastic.

Let us consider now a case when the energy of the free electron is higher than the difference of energies of the first excited and ground states of oscillator.

In this case as a result of collision the oscillator undergoes a transition to a state with higher energy than energy of the first excited state. In accordance with oscillator dynamics described by equation (12), as it was shown above, the oscillator undergoes a transition spontaneously to the first excited state and can remains here for a relatively long time without emitting the energy received in the collision with the electron. During the lifetime of first excited state of oscillator the free electron have enough time to fly away and it "does not gain back its energy". Thus inelastic scattering of electron takes place in this case.

Within the limits of these representations it is easy to explain a result of scattering of electron by the oscillator when the electron energy is enough for transition of oscillator into the second, third etc. excited states.

In other words, using this simple scheme we assume that free electron and oscillator interact directly only in the moment of their collision; electron and oscillator do not interact directly after collision, but the electron can be affected by a force action from radiation emitted by oscillator.

Let us formulate a mathematical model corresponding to these representations. It is obvious, the process under consideration is one-dimensional: motion of both electron and oscillator occurs along the same axis.

The oscillator is described by equation (12) while the energy of free electron is changed due to absorption of the energy emitted by the oscillator.

Considering approximately that the rate of energy absorption by the electron is in proportion to the rate of energy emitted by the oscillator and taking into account (4) we obtain

$$\frac{dE_e}{dt} = \zeta(t) f(q, \dot{q}, E) \tag{17}$$

where $E_e$ is electron energy; the function $f(q, \dot{q}, E)$ is defined by expression (7); factor $\zeta(t)$ takes into account a finite time of interaction of electron with radiation of oscillator.

For example, $\zeta(t) = \exp(-\beta t)$, where $\beta = const$ or stepwise function: $\zeta(t) = 1$ at $t \le t_{\text{int}}$ and $\zeta(t) = 0$ at $t > t_{\text{int}}$ can be used in the model.

In non-dimensional variables (10) the equation (17), (7) has the form

$$\frac{dE_e}{dt} = \zeta(t) \alpha \dot{q}^2 \left(\dot{q}^2 + q^2 - 1\right) \cos^2\left(\frac{\pi}{2}(\dot{q}^2 + q^2)\right) \tag{18}$$

Non-dimensional velocity of electron is $V_e = \sqrt{2E_e}$.

We assume here that the electron and oscillator have identical mass. This assumption can be explained by the fact that in real Franck-Hertz experiment the collision occurs between free electron and an electron which is inside atom.

Let us describe briefly the calculation method.

It is assumed that the oscillator exists in the ground state before collision with electron.

This means that the coordinate and velocity of the oscillator are described by the classical expressions

$$q = \sin t,\ \dot{q} = \cos t \tag{19}$$

Before collision the free electron has a velocity $V_e^0 > 0$ and energy $E_0 = \frac{1}{2}(V_e^0)^2$.

Since the process of collision of free electron with oscillator is random the moment of collision is selected randomly from the range $[0, 2\pi]$. In this case the velocity and energy of the oscillator before the collision are defined by expressions (19).

It is assumed that the coordinate of oscillator is not changed during the collision, while the velocities of the oscillator and the free electron can be calculated immediately after the collision by using the formulas of classical mechanics for particles elastic collision

$$\begin{aligned} V_0 &= \frac{1}{2}\left(\dot{q}_- + V_e^0\right) + \frac{1}{2}\left|\dot{q}_- - V_e^0\right| \\ V_e &= \frac{1}{2}\left(\dot{q}_- + V_e^0\right) - \frac{1}{2}\left|\dot{q}_- - V_e^0\right| \end{aligned} \tag{20}$$

where $\dot{q}_-$ is the velocity of the oscillator immediately before the collision.

Velocities of the electron and the oscillator immediately after their collision (20) are the initial conditions for solution of equations (12) and (18). Instant of time $t = 0$ corresponds to collision.

It is assumed in further calculations $\alpha$ =10 and the stepwise function $\zeta(t)$ at $t_{\text{int}}$ =200 was used.

Results of calculations for a single free electron are illustrated in Fig.15.

The dependencies of energies of both electron and oscillator after scattering on initial energy of electron are shown in Fig. 15.

A result of collision of a single electron with the oscillator is always random due to the random conditions of the oscillator before the collision and the random nature of transitions of oscillator from excited state to the lower energy level (see the results above). For this reason Fig. 15 shows one of a random realization of the process. For example (see Fig. 15) at some initial energy of electron the oscillator undergoes a transition to an excited state but due to instability of the excited state the oscillator has enough time to undergo a transition to the ground state and to transfer energy of its radiation to electron. As a result the scattering of the electron is elastic in this case while the scattering was inelastic for other neighbor energies of free electron.

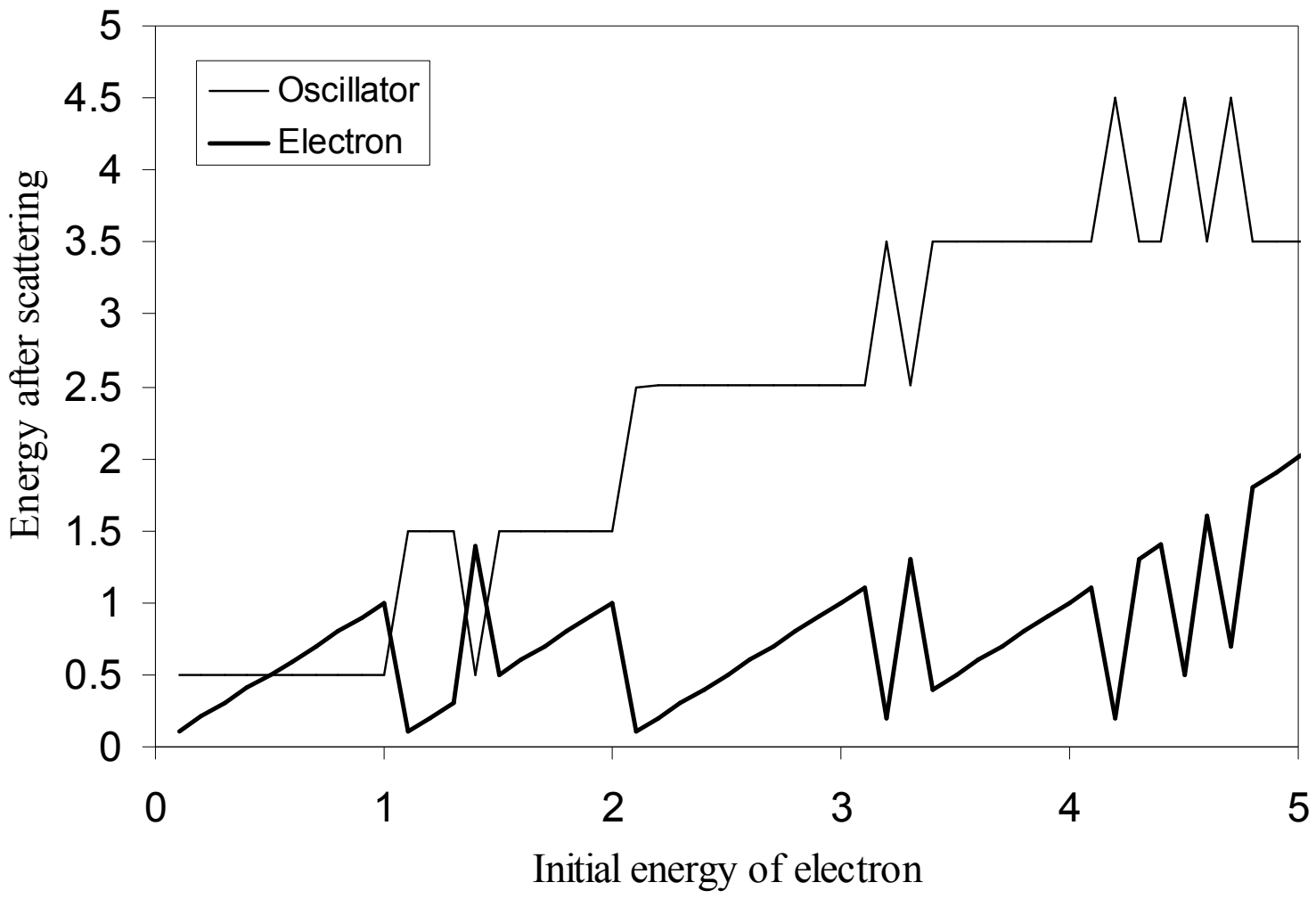

FIG. 15. Dependencies of energy of electron and oscillator after scattering on initial energy of electron

In real Franck-Hertz experiment we deal with not single electron but with electron beams where an averaging of random factors occurs.

To model the Franck-Hertz experiment it is necessary to calculate a large number of collisions of electrons with the oscillator and to average the results. 200 numerical experiments were carried out in this work for each initial electron energy with the random conditions of oscillator; the results of calculations were averaged. The dependences of mean velocity and mean energy of electrons after scattering on initial electron energy obtained in the calculations are shown on Fig.16. The mean velocity of electron in these calculations is similar to the current measured in Franck-Hertz experiment.

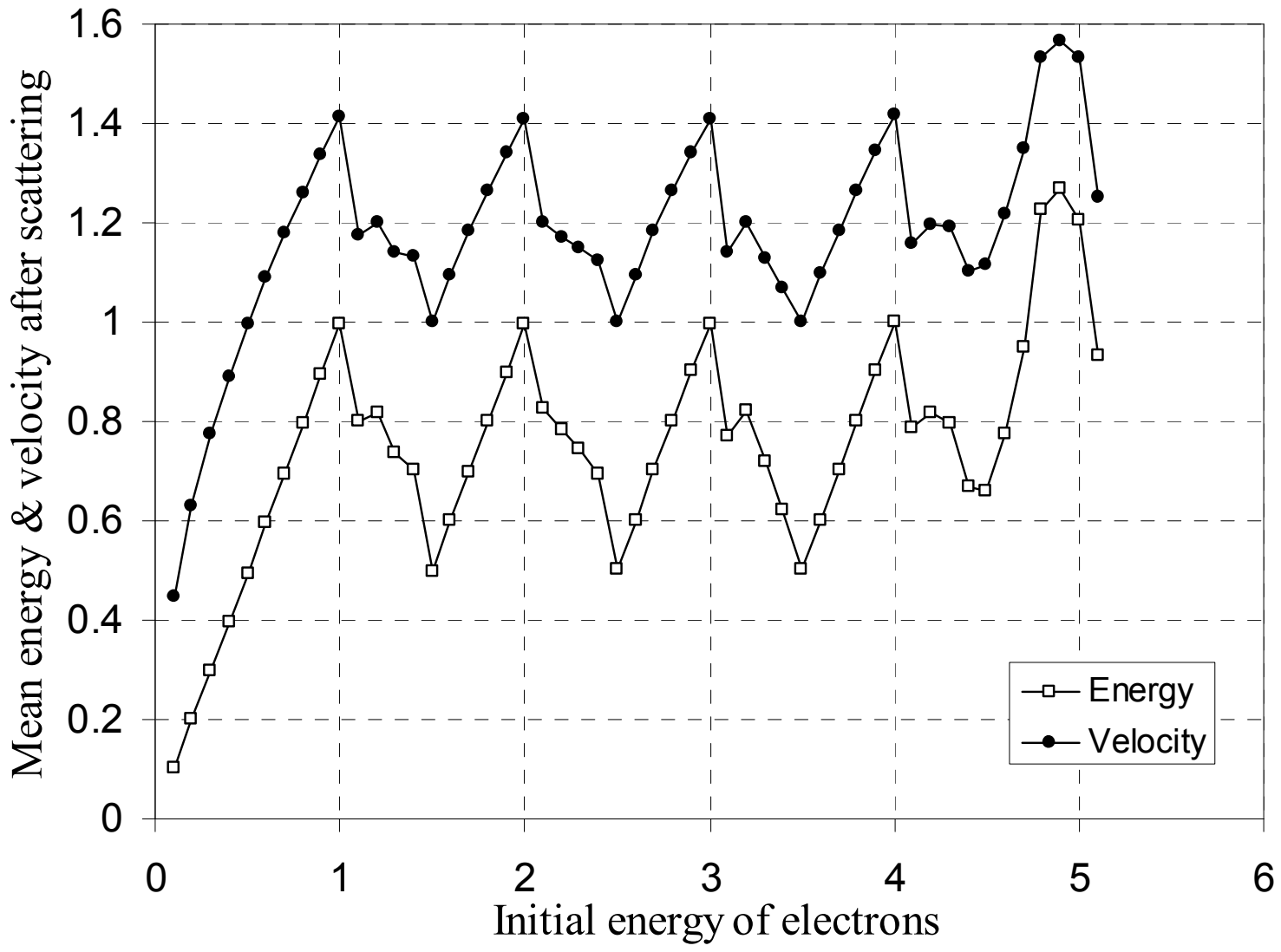

FIG. 16. Dependencies of mean energy and mean velocity of electrons after scattering on initial energy of electrons; $\alpha = 10$

The character of the curves in Fig. 16 is in qualitative agreement with the experimental data obtained from the Franck-Hertz experiment with different gases. One sees, if the energy of free electron less than the difference of energies between the first excited state and the ground state, $\hbar\omega$, the scattering of the electron by oscillator is elastic. If the energy of electron higher than $\hbar\omega$, the scattering is inelastic.

Peaks magnitude and their locations are explained by the fact that distances between all neighboring energy levels in the oscillator are the same and equal to $\hbar\omega$.

The distances between different neighbor levels in atoms are different. Therefore the peaks magnitude and distances between them are usually different too in real experiments.

The local non-monotonicity observed in calculated dependences on Fig. 16 are connected with the random character of the process of collision and the relatively small number of electrons (200) which were used for averaging.

It has to be noted that obtained results of calculations correspond to a low frequency of electron-oscillator collisions where the oscillator has enough time between consecutive collisions to come back into the ground state. As we have shown above the lifetime of the oscillator drastically decreases with increasing of quantum number $n$ (Fig. 13). This means that the time of interaction $t_{\text{int}}=200$ used in calculations may be not enough for oscillator to come back into the ground state in particular if free electron has a high initial energy. It is clearly shown at Fig. 16: for non-dimensional initial energy of electron $E_0 > 4$ the results are essentially different than those for $E_0 < 4$. It can be expected that for high frequency of electron-oscillator collisions the results will depend on the ratio of lifetime of excited states and a time period between consecutive collisions of free electrons in electron beam with oscillator. This case corresponds to intense electron beam in the Franck-Hertz experiments where the multiple collisions can occur. The model under consideration shows a way for solving of this problem.

## VI. CONCLUSION

We have shown that it is possible to construct a classical dynamical system which possesses a specific set of properties which are characteristic for real quantum objects.

Within the limits of the model under consideration it was possible to give a classical explanation of the Franck-Hertz experiment. It is hoped that this will allow a deeper understanding of this complicated process.

It has to be noted that the proposed model does not pretend to give an alternative description of quantum systems because it does not take into account the particle-wave duality which is a fundamental property of quantum objects.

We think that a development of similar classical models of dynamical systems with «quantum behavior» can have the following purposes.

1. To show that it is possible to construct the systems within the limits of the classical mechanics which will possess the properties similar to some of the properties of the real quantum systems.
2. Such models may be useful for semi-classical molecular dynamics simulations of real systems with taking into account the spontaneous and stimulated transitions, interactions of atoms with radiation (absorption and scattering of photons) etc. The above developed classical model of the Franck-Hertz experiment gives some optimism in this direction. Within the limits of classical dynamics such models allow us to follow in time a behavior of separate atom during the emission or absorption of energy. Such observations cannot be made by quantum mechanics which predicts only the final result of the process in the form of probabilities of different events. It is expected that for some systems a choice of parameters of the model allows obtaining not only qualitative, but also quantitative (in statistical sense) agreements of the results of the classical dynamical model with the results of exact quantum calculations as per Schrödinger equation.
3. Such models may be useful for semi-classical calculations of Rydberg atoms.